%% file: T0-main.tex
\documentclass[modern,twocolumn]{aastex63}
\begin{document}

\title{A Case study of light pollution in France}
\shorttitle{Light pollution changes in France}
\shortauthors{Aksaker et al.}
    \received{receipt date}
    \revised{revision date}
    \accepted{acceptance date}
    \published{published date}
    \submitjournal{Ap. Sp.\&Sci.}

\author{N. Aksaker}
    \affiliation{Adana Organised Industrial Zones Vocational School of Technical Science, University of Çukurova, 01410, Adana, Turkey.}
    \affiliation{Space Science and Solar Energy Research and Application Center (UZAYMER), University of Çukurova, 01330, Adana, Turkey.}
    \correspondingauthor{N. Aksaker}
    \email{naksaker@cu.edu.tr}
\author{S.K. Yerli}
    \affiliation{Department of Physics, Orta Doğu Teknik Üniversitesi, 06800, Ankara, Turkey.}
\author{Z. Kurt}
    \affiliation{Remote Sensing and Geographic Information System, University of Cukurova, 01330, Adana, Turkey}
    \affiliation{Space Science and Solar Energy Research and Application Center (UZAYMER), University of Çukurova, 01330, Adana, Turkey.}
\author{M. Bayazit}
    \affiliation{Remote Sensing and Geographic Information System, University of Cukurova, 01330, Adana, Turkey}
\author{A. Aktay}
    \affiliation{Turkish State Meteorological Service, Regional Forecast Center, 01360, Adana, Turkey}
\author{M.A. Erdoğan}
    \affiliation{Landscape Architecture Department, Faculty of Architecture, Hatay Mustafa Kemal University, 31060 Hatay, Turkey.}

\begin{abstract}
In this study, we investigate the effects of light pollution in France using GIS data which was first used in \cite{2020MNRAS.493.1204A} (so called astroGIS database - \url{astrogis.org}). A subset of Artificial Light layer of astroGIS database has been adapted for years between January 2012 and December 2019. During 2019, half a million of lumen has been released into space from the total surface area of France. Annual light pollution in France has a decreasing trend. France continues to have potential Dark Sky Park locations for example cities like Cantal, Lot, Lozère and Creuse having the lowest light pollution values. Light pollution is strongly correlated with population ($R^{2}\simeq 0.85$) and GDP ($R^{2}\simeq 0.84$). In addition, half of the observatories remain under the light polluted areas.
\end{abstract}

\keywords{Light Pollution}

\section{Introduction}

The world surface continuously perturbed by the humanity and this can be seen from space at night.
This effect is identified as Artificial Light at Night, namely AL or simply \textit{light pollution}
\citep{2020MNRAS.493.2463C, 2020AAS...23540103M, 2020ERCom...2....2S}.
Light pollution is one of the most damaging effects for astronomy. At the moment, one-third of humanity is not aware of the Milky Way because they \textbf{cannot} see it. 
Moreover, 80\% of the world population lives in light polluted regions \citep{falchi2016}. 

Astronomical objects simply disappear in the night sky when telescopes point to them in observatories which are affected by the cumulative light above the large cities \citep{Gronkowski2017}.
Horizontal beams of artificial light emerging from the large cities have great harmful effects on the observatories centered on a circle with radius around 50--100 km \citep{doi:10.1063/1.3273014}.
The Technical Office for the Protection of Sky Quality (OTPC; part of IAC in Spain) provides advice on the implementation of the ``Sky Law'' (Number: 31/1988) which protects the astronomical quality of observatories in the Canaries from light, radioelectrical and atmospheric pollutions, and aviation routes.
France, as an EU member state, acted further and introduced a new regulation on light pollution to prevent emission of light in outdoor spaces\footnote{France Legislation: 28.12.2018/17}.

On the other hand, the natural relationship between artificial light and human activity such as lightening, population, national domestic product suggest strong correlations with the socioeconomic factor \citep{2017SciA....3E1528K, 2018PhyA..492.1088O}.

In this study, we investigate the effects of light pollution in France using the astroGIS database \cite{2020MNRAS.493.1204A}.
Details of Artificial Light (AL) layer in astroGIS database can be accessed online%
\footnote{\url{astrogis.org}}. A general view of the adapted dataset from astroGIS database is given in Fig. \ref{F:france}.

\section{GIS and Nighttime Dataset}
\label{sec:data}
Geographical Information System (GIS) is a robust, easy to use and, time and cost efficient technique for geo-spatial analysis.
The GIS data can also be produced by using remote sensing techniques from satellite imagery.
Our astroGIS database \citep{2020MNRAS.493.1204A} contains GIS dataset with several layers.
The demographic dataset of this study contains digitized data from recent GADM\footnote{\url{gadm.org}} (Database of Global Administrative Areas) and GDP\footnote{\url{stats.oecd.org}} (Gross Domestic Product) for 2015.
France has 96 cities and, their boundaries and total surface area were digitized from GADM dataset (see Table \ref{T:cities}).
The Table \ref{T:cities} also contains official municipality population data\footnote{\url{data.gouv.fr}}, dated on 1 January 2017.
The summary of demographic data can be viewed in the left panel of Fig. \ref{F:france}. 

The nighttime data are taken from the Visible Infrared Imaging Radiometer Suite (VIIRS) instrument on board SUOMI-NPP satellite in the Day Night Band (DNB) which corresponds to visible spectrum.
The radiometric resolution is up to 14 bits in between visible and red ($0.5-0.9 \mu$m) which gives a minimum of approximately $2 \times 10^{-9}$ W cm$^{-2}$ sr$^{-1}$ radiation counts \citep{nurbandi2016}.
The GIS dataset containing daily produced, nighttime images cover the earth's surface from 75 N to 65 S latitudes with 15$\arcsec$ grid size in a GEOTIFF format corresponding to a spatial resolution of 463 m per pixel.
The nighttime data filtered out from non-artificial light sources (e.g. lightning, fishing boats, clouds etc.) by \citet{elvidge2017} and, monthly and yearly averaged images are stored in a publicly accessible database \footnote{\url{eogdata.mines.edu}}.
VIIRS data is presented to end user in six layouts where France's surface area is found under ``Tile 2''.
The dataset contains 558 images for the whole time span from April 2012 to December 2019.
Each image's storage size is approximately 1.6 GB, therefore, the dataset mounts up to 1 TB.
As an example, overall view of the nighttime data for December 2019 is given in the right panel of Fig. \ref{F:france}.

\section{Analysis of the data}
\label{sec:analysis}
Monthly averaged nighttime data from astroGIS data\-base contains 93 images from April 2012 to December 2019.
Surface area of France was extracted using digitized GADM boundaries.
Furthermore, each city in France have to be extracted from the same dataset.
Using a pre-filtering algorithm in Python, above $3 \sigma$ values with respect to average light pollution value over the whole time span were excluded for each pixel.
A model in Zonal Statistics tool of ArcGIS Desktop 10.4.1 has to be created to process 96 cities in total to calculate pixel averages within each city boundary.
The filtered-out light pollution data is produced by using monthly nighttime data for each city.
Yearly averages were also calculated from the monthly data (Table \ref{T:result}).
In the table, a linear regression fit applied to yearly averages (column `L.R.') along with goodness of fits (R$^2$).

Earth Observation Group (EOG) updated VIIRS sensor calibration for Airglow after the 2017 \citep{2019ITGRS..57.9602U}. 
\cite{2020Sens...10.1964C} produced and published \cite{2019ITGRS..57.9602U} a mask for this new calibration.
This airglow correction has been applied to whole dataset and used through out in all stages.

Using standard equations (see e.g. \cite{2020MNRAS.496L.138K}) radiance energy (nW cm$^{-2}$ sr$^{-1}$) transformed into luminous flux values (lm).
Therefore total flux of a city can be calculated by summing the flux over all the pixel of the city ($Area\times Luminosity$).
Detection limit of VIIRS DNB detector is 3 nW cm$^{-2}$ sr$^{-1}$ \citep{2013JGRD..11812705L}.
Since values below this limit create noise on the flux, the limit is used as a filter and dataset is cleared out from this noise before applying to any further calculations.

In total 107 observatories have been noted through out France (Table \ref{T:obs}).
Their locations are taken as point sources due to spatial resolution of VIIRS sensor (463 m per pixel).
52 observatories show no AL variations or stayed below the sensor limit of 3 nW cm$^{-2}$ sr$^{-1}$ and therefore, they are marked as dashes.
Due to their point source nature, observatories can easily be judged according to their sky brightness if AL values are converted to well-known magnitude scale.
One of the easiest way of obtaining this quality is to use a Sky Quality Meter (SQM) where it measures the light it detects in units of mag/arcsec$^{2}$ (mpsas).
Our AL dataset for each observatory is converted to SQM values using the following equation \citep{2020NatSR..10.7829S}:
\begin{equation}
20.0 - 1.9 \log(\textrm{AL})
\end{equation}
where AL is the VIIRS DNB value in nW cm$^{-2}$ sr$^{-1}$ units.
See also Table \ref{T:result} for the definitions of other columns.
The sense of ``AL for observatories of France'' is shown in Fig. \ref{F:obs}.
In this contour map graph, light (dark) colors represents higher (lower) AL values in SQM scale.
All of the observatories in Table \ref{T:obs} are overlayed on the figure with blue filled circles.
Note that most of the observatories (even the ones high in altitude) are hidden behind or surrounded by the AL dominated cities.

Note also that City Emission Function (CEF) which is defined as the angular emission from an entire city as a light source \citep{2019PNAS..116.7712K} could have a major effect on the satellite signal due to the fact that AL coming from surface sources might have seasonal and/or wavelength dependence.
Since VIIRS data is integrated over whole spectral response of the DNB, this CEF effect might have to be taken into account if the source (cities or regions) contains mixed type of lighting systems (e.g LEDs, High Pressure Sodium Lamps etc.).
However, in this work we focus on total energy from the surface therefore we can simply neglect the wavelength dependence and potential seasonal changes.

\section{Results and Discussions}
\label{sec:result}
We investigated the light pollution dataset (running from January 2012 to December 2019) for France created from our earlier astroGIS database \citep{2020MNRAS.493.1204A}.
The analysis of the France dataset can be concluded with the following outcomes:
\begin{itemize}

    \item During 2019 total energy released into space from the total surface area of France was approximately half a million lumen. This value for 2019 shows a relatively decreasing trend which is very close to 2012 level.

    \item During the time span of the data set, almost all cities shows a decreasing AL trend (Table \ref{T:result}).
    However, as can be seen in Fig. \ref{F:dist}, cities grouped as ``typical'' show a linear steady trend before 2017, and after 2017, which is when an important calibration update has been applied to VIIRS satellite data, disruption of steadiness has been observed over almost all cities causing a negative, decreasing trend in general (see the linear regression column of Table \ref{T:result}).
    Therefore, to be able to describe a good interpretation of the trend, one has to gather more data with the new calibrated VIIRS detector.

    The enactment of artificial light legislation might control this huge energy release, however, to have a sense of whether it works or not one has to accumulate more data than recorded.

    \item Through the time span of the dataset, strongly light polluted cities tend to show slight improvements, however, the cities marked as ``under luminous'' show almost a constant trend whereas ``over-luminous'' and ``luminous'' cities stay constant for the whole period, regardless of the calibration change occurred in 2017 (see Fig. \ref{F:dist}).

    \item When meteorological and astronomical parameters (e.g.\ elevation and cloud coverage) excluded, France has dozens of potential \textbf{Dark Sky Park} locations where Cantal, Lot, Lozère and Creuse can be counted as the darkest among the others. Moreover, Darker regions can easily seen on Fig. \ref{F:obs}.

    \item When the light pollution distribution over the whole country considered for 2019, geographical ``points'' (i.e. pixels in our dataset) Creuse and Nord have the minimum (10.5) and maximum (91,468.36) values in lumen, respectively.

    \item Correlating human activity (e.g. electric consumption) to the energy escaped to space is given as an example in \cite{2014RemS....6.1705S} for China.
    We also found a strong correlation between population and light pollution with 0.85 confidence in Table \ref{T:cities} (see also Fig. \ref{F:pop_flux}). A similar good correlation can also be observed for GDP with 0.84 confidence.
    
    \item Half of the observatories remain under the detection limit of the VIIRS satellite which are marked with dashes in Table \ref{T:obs} and Fig \ref{F:obs}.
    Therefore, it is easy to conclude that the other half of the observatories are located in light polluted areas. 
    
    \item In conclusion, it is too early to justify whether the improvements observed in the dataset are due to the enactment of the legislation or not.
\end{itemize}

\section*{Acknowledgements}
This research was supported by the Scientific and Technological Research Council of Turkey (TÜBİTAK) through project number 117F309. This research was also supported by the Çukurova University Research Fund through project number FYL-2019-11770.
We also would like to thank to the reviewer for his/her positive and constructive comments. 

\paragraph{Compliance with ethical standards} The authors declare that they have no potential conflict and will abide by the ethical standards of this journal.

\bibliographystyle{aasjournal}
\bibliography{T3-main}

\input{T1-Fig.tex}
\input{T2-tab-1.tex}
\input{T2-tab-2.tex}
\input{T2-tab-3.tex}

\end{document}

%% file: T1-Fig.tex
\begin{figure*}
    \centering
	\includegraphics[width=0.49\textwidth]{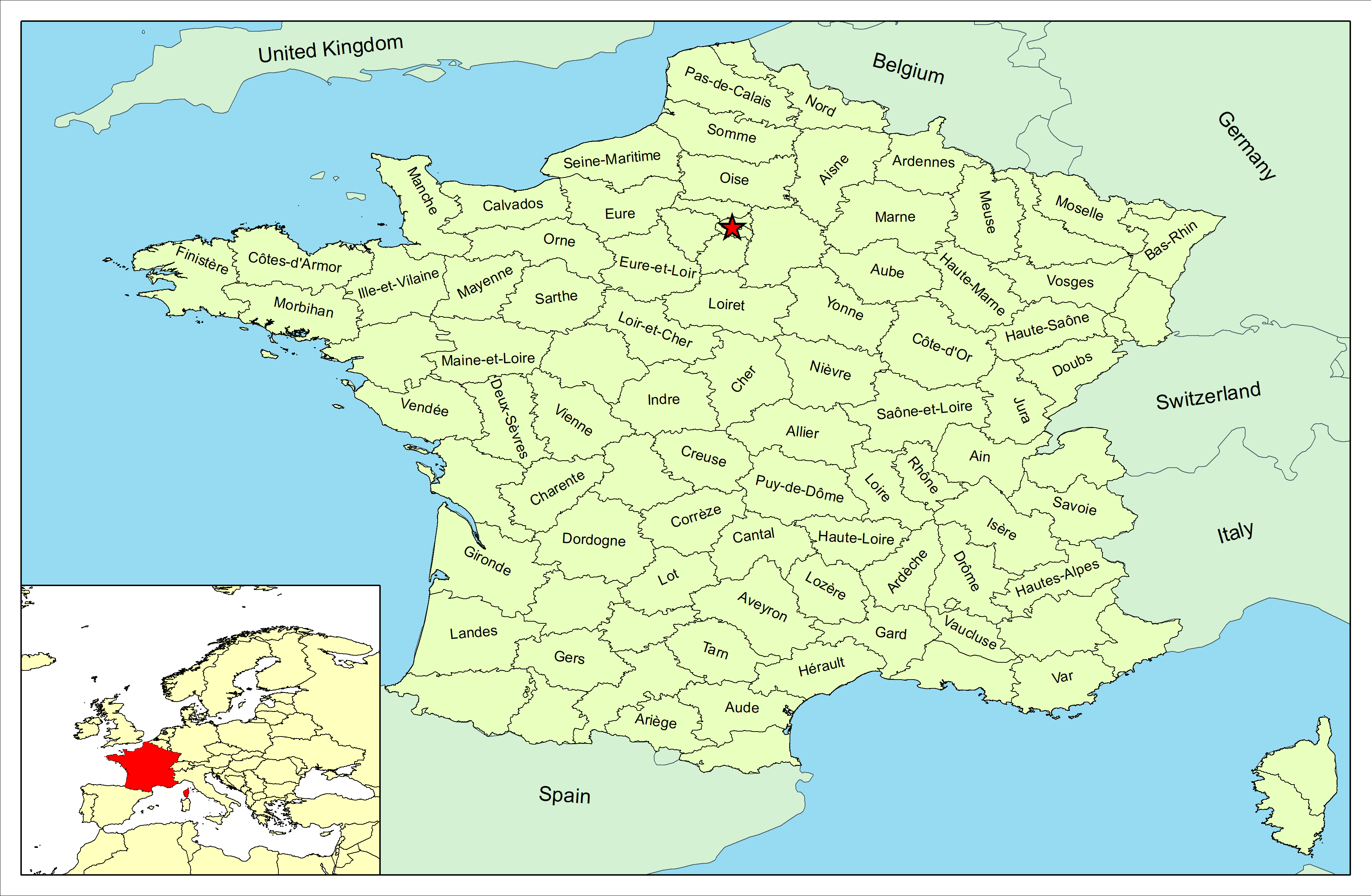}
	\includegraphics[width=0.49\textwidth]{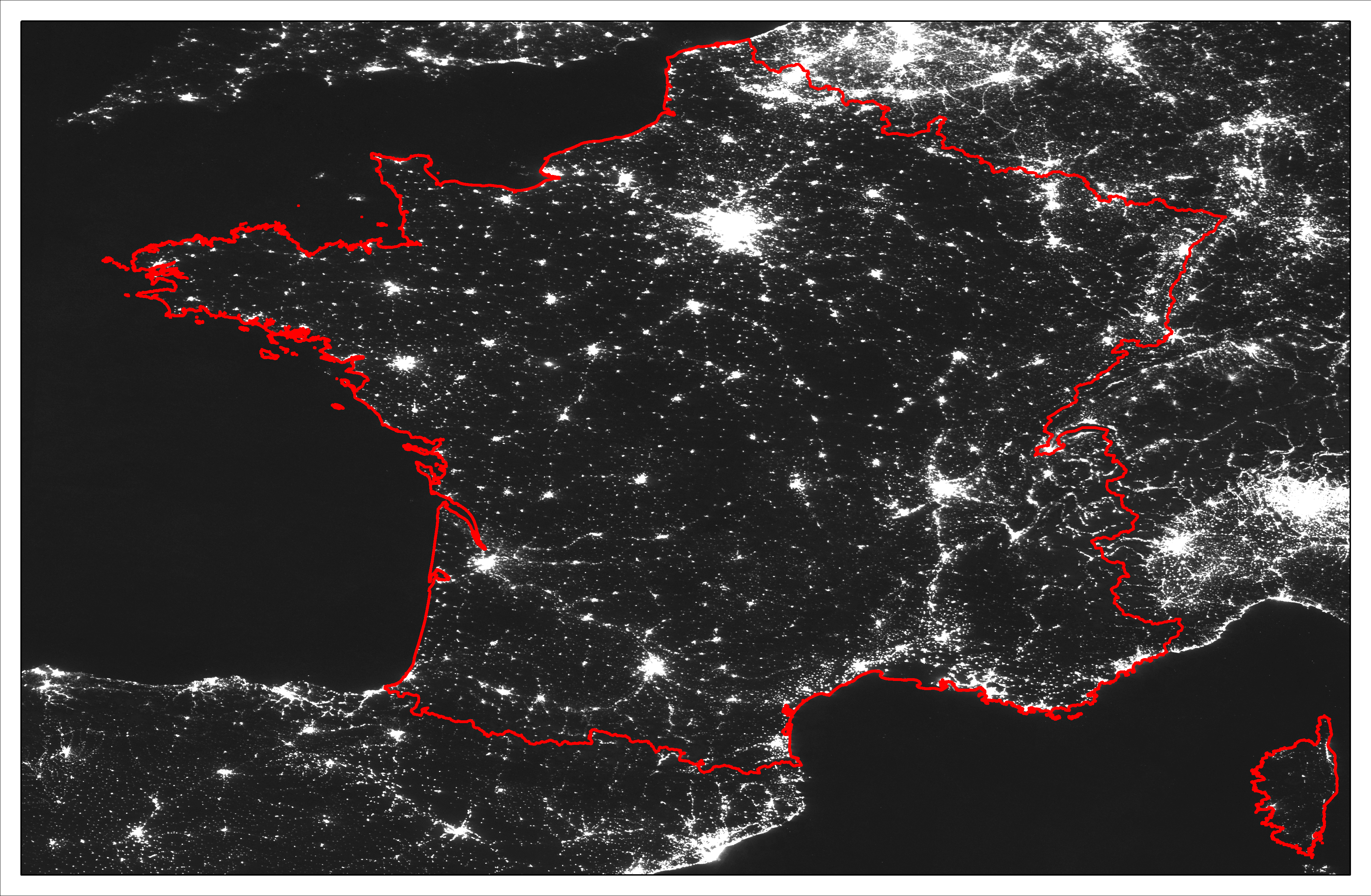}
	\caption{%
	Left Panel: Demographic map of France showing the name and boundaries of each city.
	Names not fitting to their boundaries are left as unnamed.
	The red star marks the capital city, Paris.
	Boundary of neighbouring countries are also drawn with no other details.
	Right Panel: Artificial Light (AL) distribution of France for December 2019.
	AL seen from space is colored as white.
	As expected, Paris region and other heavily populated major cities dominate the AL distribution.
	Note also that geographically less populated regions, for example, rural areas, mountains, lakes etc. are colored with black.}
	\label{F:france}
\end{figure*}

\begin{figure}
    \centering
	\includegraphics[width=\columnwidth]{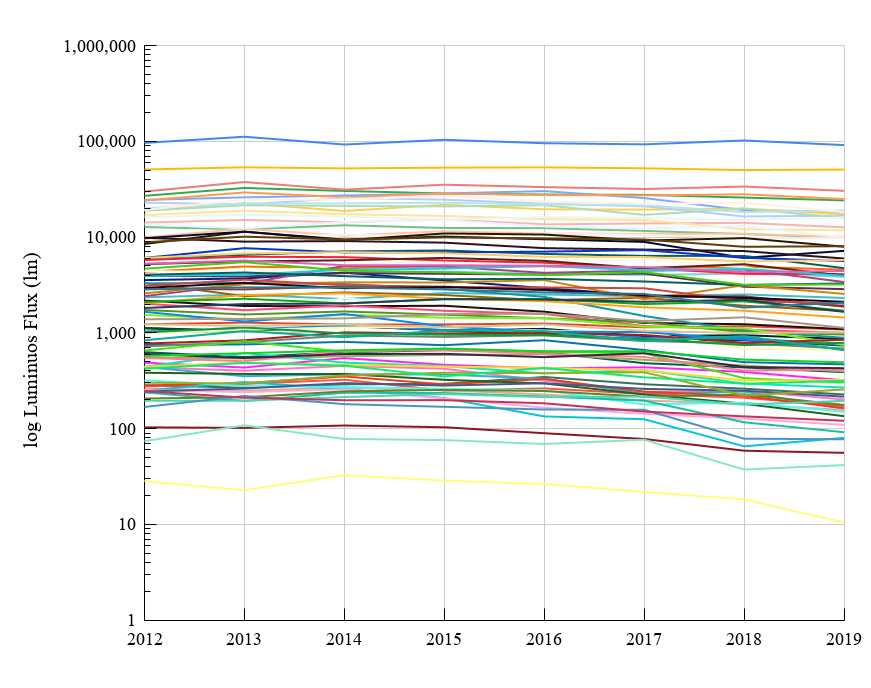}
	\caption{%
	Luminous flux of Artificial Light (AL) averaged annually for all cities in between 2012--2019.
	Even though a general ``steady variation'' exists in the graph, to be able to deduce a representative trend though out country cities can easily be merged into four groups: over-luminous ($<$50,000), luminous ($<$10,000), typical (100 -- 10,000), under-luminous ($<$100).
	It is common to mark over- and under-luminous cities as outliers.
	Therefore AL of cities marked as ``typical'', can be summarized as (a) slightly increasing starting from 2012, (b) slightly decreasing after 2017, (c) in between 2013-2017 remaining steady.%
}
	\label{F:dist}
\end{figure}


\begin{figure*}
    \centering
	\includegraphics[width=0.49\textwidth]{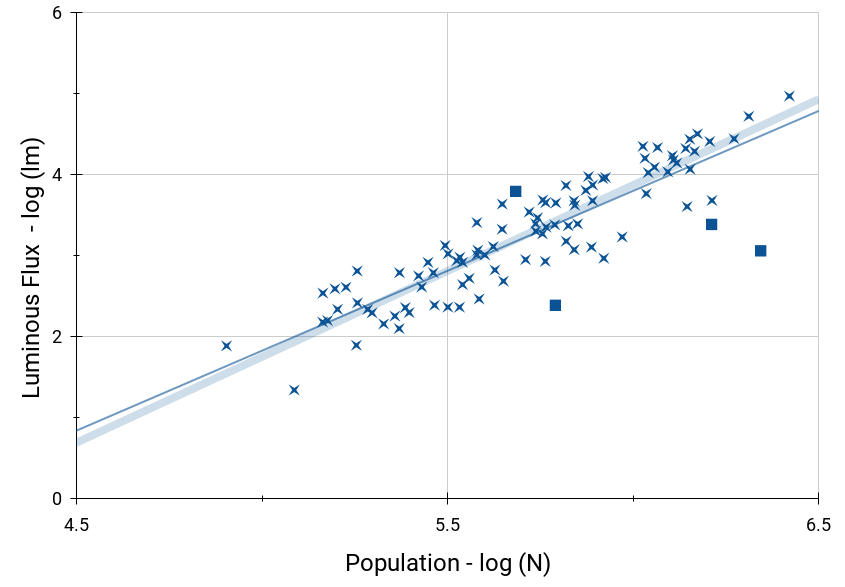}
	\includegraphics[width=0.49\textwidth]{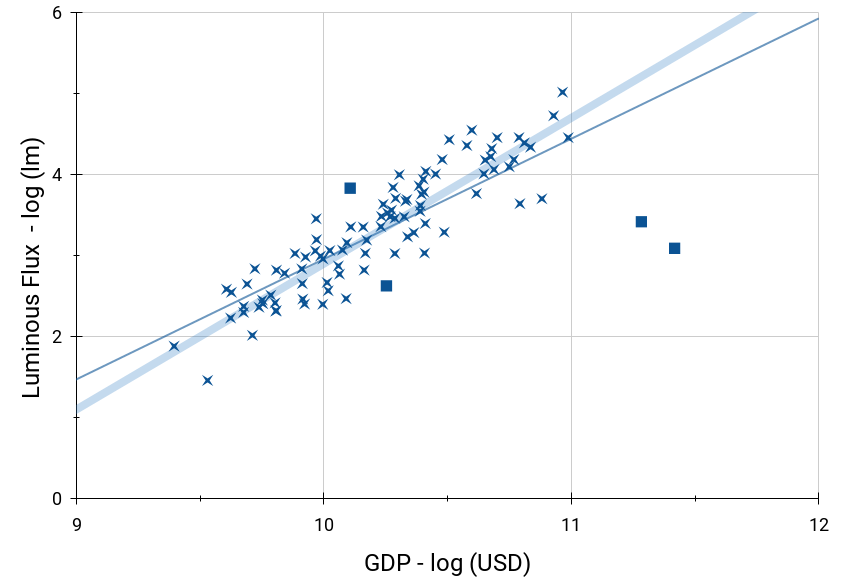}
	\caption{%
	Left Panel:
	Total flux versus population for the year of 2017 in logarithmic scale.
	Thick solid line represents the linear regression \textbf{($y=2.12x-8.85, R^2\sim 0.85$)}.
	Right Panel:
	Total flux versus GDP (USD) in logarithm scale.
	Thick solid line represents the linear regression \textbf{($y=1.80x-15.1, R^2\sim 0.84$)}.
	Note that, some of the cities can easily be marked as outliers due to their excessive GDP or Population values, namely Pyrénées-Orientales, Hauts-de-Seine, Paris, Côtes-d'Armor (ordered from highest to lowest).
	Therefore, 4 cites (solid squares) are taken out from the linear regressions.
	Thin solid lines represent the linear regressions when outliers were included.%
	}
	\label{F:pop_flux}
\end{figure*}

\begin{figure*}
    \centering
	\includegraphics[width=\textwidth]{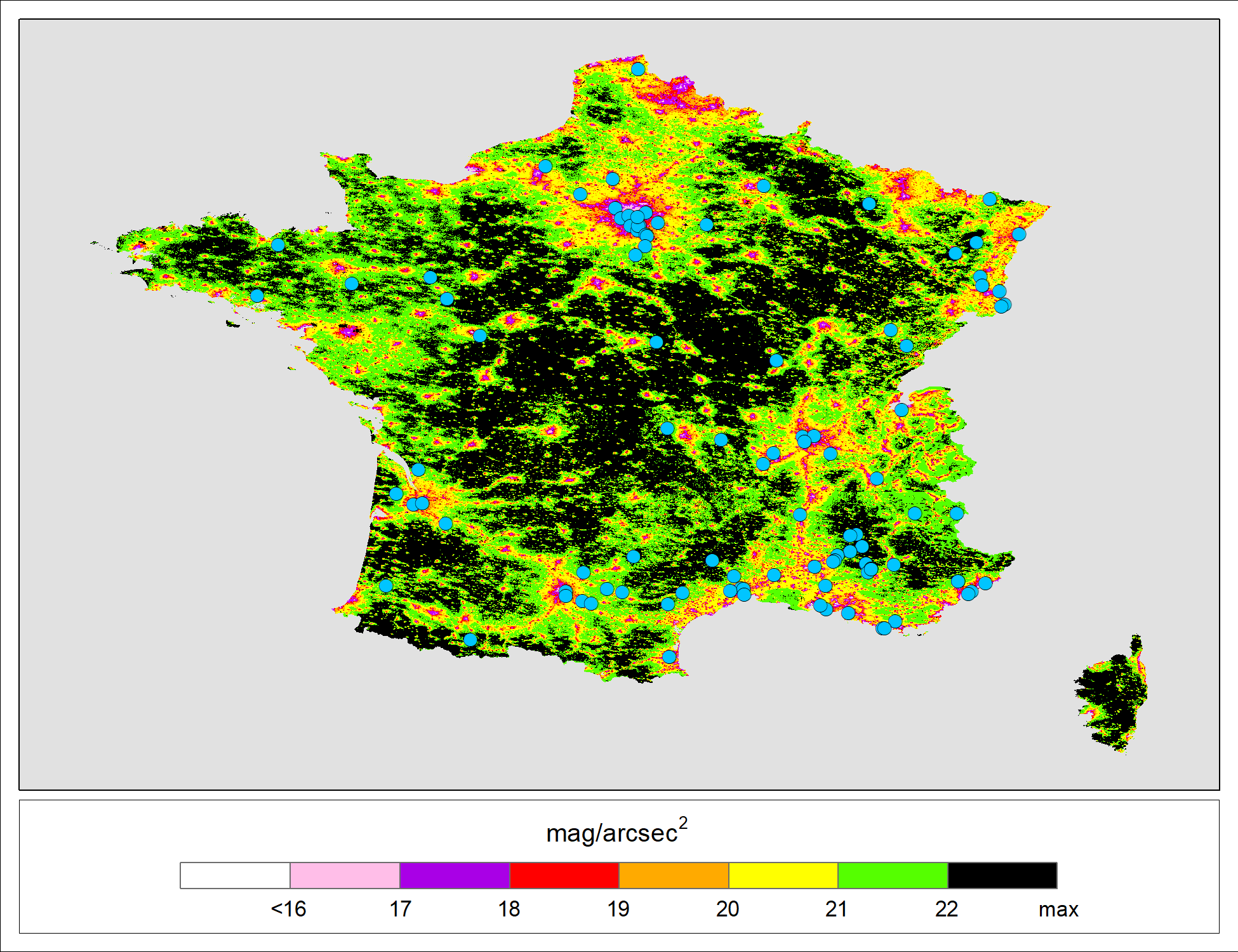}
	\caption{%
    Observatories of France (blues filled circles) overlaid on to the yearly (2019) Artificial Light (AL) of the country.
    The colors represents AL in unit of mag/arcsec$^{2}$.
   Note that most of the observatories are severely effected from the AL.%
    }
	\label{F:obs}
\end{figure*}

%% file: T2-tab-1.tex
\begin{table*}
    \caption{%
    Limited number of demographic values for all cities in France.
    Surface area is in square km, GDP is in billions of USD and population is for 2017.
    See section \ref{sec:data} for the discussion on relating surface to observed pixel coverage.%
    }
    \begin{center}
    \begin{footnotesize}
    \begin{tabular}{@{}lrrr@{\hspace*{3em}}lrrr@{}}
    \hline 
    City   & Area & Population & GDP &
    City   & Area & Population & GDP\\
    \hline
Ain	&	5.775	&	631,877	&	19.0	&	Indre-et-Loire	&	6.149	&	604,966	&	21.7	\\[-2pt]
Aisne	&	7.426	&	538,659	&	14.4	&	Isère	&	7.865	&	1,251,060	&	47.2	\\[-2pt]
Allier	&	7.366	&	341,613	&	10.3	&	Jura	&	5.041	&	260,587	&	8.2	\\[-2pt]
Alpes-de-Haute-Provence	&	7.026	&	161,799	&	4.9	&	Landes	&	9.352	&	403,234	&	14.9	\\[-2pt]
Alpes-Maritimes	&	4.321	&	1,082,440	&	44.3	&	Loir-et-Cher	&	6.41	&	333,050	&	10.4	\\[-2pt]
Ardèche	&	5.56	&	324,209	&	8.4	&	Loire	&	4.798	&	759,411	&	24.8	\\[-2pt]
Ardennes	&	5.249	&	277,752	&	7.6	&	Loire-Atlantique	&	6.893	&	1,365,227	&	56.1	\\[-2pt]
Ariège	&	4.93	&	152,499	&	4.0	&	Loiret	&	6.803	&	673,349	&	25.4	\\[-2pt]
Aube	&	6.021	&	309,056	&	9.2	&	Lot	&	5.223	&	173,400	&	5.1	\\[-2pt]
Aude	&	6.268	&	366,957	&	9.3	&	Lot-et-Garonne	&	5.382	&	333,417	&	9.7	\\[-2pt]
Aveyron	&	8.771	&	279,169	&	8.1	&	Lozère	&	5.172	&	76,309	&	2.5	\\[-2pt]
Bas-Rhin	&	4.795	&	1,116,658	&	44.9	&	Maine-et-Loire	&	7.221	&	810,186	&	25.5	\\[-2pt]
Bouches-du-Rhône	&	5.254	&	2,016,622	&	84.8	&	Manche	&	6.059	&	499,287	&	14.7	\\[-2pt]
Calvados	&	5.605	&	693,579	&	24.5	&	Marne	&	8.191	&	572,293	&	21.4	\\[-2pt]
Cantal	&	5.769	&	146,219	&	4.2	&	Mayenne	&	5.207	&	307,940	&	9.9	\\[-2pt]
Charente	&	5.964	&	353,613	&	11.6	&	Meurthe-et-Moselle	&	5.281	&	734,403	&	24.2	\\[-2pt]
Charente-Maritime	&	6.934	&	639,938	&	19.4	&	Meuse	&	6.236	&	190,626	&	5.7	\\[-2pt]
Cher	&	7.296	&	308,992	&	9.3	&	Morbihan	&	6.874	&	744,813	&	23.1	\\[-2pt]
Corrèze	&	5.891	&	241,871	&	8.4	&	Moselle	&	6.252	&	1,044,486	&	32.1	\\[-2pt]
Corse-du-Sud	&	4.049	&	152,730	&	5.5	&	Nièvre	&	6.862	&	211,747	&	6.4	\\[-2pt]
Côte-d'Or	&	8.789	&	533,147	&	21.1	&	Nord	&	5.765	&	2,605,238	&	92.2	\\[-2pt]
Côtes-d'Armor	&	7.029	&	598,357	&	17.9	&	Oise	&	5.894	&	821,552	&	25.8	\\[-2pt]
Creuse	&	5.589	&	120,365	&	3.4	&	Orne	&	6.143	&	286,618	&	8.2	\\[-2pt]
Deux-Sèvres	&	6.029	&	374,435	&	12.3	&	Paris	&	105	&	2,206,488	&	261.3	\\[-2pt]
Dordogne	&	9.212	&	415,417	&	11.4	&	Pas-de-Calais	&	6.73	&	1,472,648	&	39.6	\\[-2pt]
Doubs	&	5.244	&	536,959	&	18.5	&	Puy-de-Dôme	&	7.999	&	647,501	&	25.7	\\[-2pt]
Drôme	&	6.554	&	504,637	&	18.8	&	Pyrénées-Atlantiques	&	7.699	&	670,032	&	24.5	\\[-2pt]
Essonne	&	1.818	&	1,276,233	&	58.6	&	Pyrénées-Orientales	&	4.131	&	471,038	&	12.8	\\[-2pt]
Eure	&	6.037	&	601,948	&	17.1	&	Rhône	&	3.253	&	1,821,995	&	97.2	\\[-2pt]
Eure-et-Loir	&	5.928	&	434,035	&	12.8	&	Saône-et-Loire	&	8.599	&	555,408	&	17.0	\\[-2pt]
Finistère	&	6.848	&	907,796	&	30.6	&	Sarthe	&	6.237	&	568,445	&	19.4	\\[-2pt]
Gard	&	5.877	&	738,189	&	20.2	&	Savoie	&	6.297	&	428,204	&	17.4	\\[-2pt]
Gers	&	6.303	&	190,932	&	4.7	&	Seine-et-Marne	&	5.925	&	1,390,121	&	50.1	\\[-2pt]
Gironde	&	10.084	&	1,548,478	&	61.5	&	Seine-Maritime	&	6.335	&	1,257,699	&	47.7	\\[-2pt]
Haut-Rhin	&	3.528	&	762,607	&	25.2	&	Seine-Saint-Denis	&	237	&	1,592,663	&	76.0	\\[-2pt]
Haute-Corse	&	4.727	&	174,553	&	5.3	&	Somme	&	6.227	&	571,879	&	18.1	\\[-2pt]
Haute-Garonne	&	6.369	&	1,335,103	&	64.6	&	Tarn	&	5.782	&	386,543	&	10.6	\\[-2pt]
Haute-Loire	&	4.998	&	227,034	&	6.4	&	Tarn-et-Garonne	&	3.729	&	255,274	&	6.4	\\[-2pt]
Haute-Marne	&	6.25	&	179,154	&	5.6	&	Territoire de Belfort	&	611	&	144,483	&	4.7	\\[-2pt]
Haute-Saône	&	5.383	&	237,706	&	6.1	&	Val-d'Oise	&	1.254	&	1,372,389	&	48.7	\\[-2pt]
Haute-Savoie	&	4.872	&	793,938	&	28.3	&	Val-de-Marne	&	246	&	1,215,390	&	62.0	\\[-2pt]
Haute-Vienne	&	5.55	&	375,795	&	12.4	&	Var	&	6.049	&	1,048,652	&	30.1	\\[-2pt]
Hautes-Alpes	&	5.718	&	140,916	&	4.2	&	Vaucluse	&	3.578	&	557,548	&	19.5	\\[-2pt]
Hautes-Pyrénées	&	4.534	&	228,582	&	6.9	&	Vendée	&	6.775	&	666,714	&	21.8	\\[-2pt]
Hauts-de-Seine	&	175	&	1,601,569	&	191.8	&	Vienne	&	7.024	&	434,887	&	14.5	\\[-2pt]
Hérault	&	6.231	&	1,120,190	&	37.8	&	Vosges	&	5.892	&	372,016	&	11.9	\\[-2pt]
Ille-et-Vilaine	&	6.854	&	1,042,884	&	41.4	&	Yonne	&	7.448	&	340,903	&	9.9	\\[-2pt]
Indre	&	6.885	&	224,200	&	6.4	&	Yvelines	&	2.304	&	1,427,291	&	68.4	\\
    \hline 
    \end{tabular}
    \end{footnotesize}
    \end{center}
\label{T:cities}
\end{table*}

%% file: T2-tab-2.tex
\begin{table*}
    \caption{%
    Annual Artificial Light at Night (ALN) for all cities of France.
    Light pollution values are in lm.
    Two different values are given for ALN: Ave-All, Ave-19 representing average of annual ALN values in between 2012-2019, for 2019, respectively.
    L.R. and $R^2$ columns are the slope of linear regression and its correlation coefficient of the regression, respectively for annual ALN values in between whole range of 2012--2019.
    See section \ref{sec:data} for the discussion on the trend of the change.
    }
    \begin{center}
    \begin{footnotesize}
    \begin{tabular}{%
        @{}             l@{~~}r@{~~}r@{~~}r@{~~}r
        @{\hspace*{1em}}l@{~~}r@{~~}r@{~~}r@{~~}r@{}}
    \hline
    City
        & \multicolumn{1}{@{}c@{}}{Ave-All}
        & \multicolumn{1}{@{}c@{}}{Ave-19}
        & \multicolumn{1}{@{}c@{}}{L.R.}
        & \multicolumn{1}{@{}c@{}}{$R^2$} &
    City
        & \multicolumn{1}{@{}c@{}}{Ave-All}
        & \multicolumn{1}{@{}c@{}}{Ave-19}
        & \multicolumn{1}{@{}c@{}}{L.R.}
        & \multicolumn{1}{@{}c@{}}{$R^2$} \\
    \hline
Ain	&	6714.20	&	5636.65	&	-115.65	&	0.31	&	Indre-et-Loire	&	4037.80	&	3404.58	&	134.55	&	0.83	\\[-2pt]
Aisne	&	2096.73	&	1708.47	&	-42.53	&	0.60	&	Isère	&	15542.58	&	11799.61	&	-931.06	&	0.69	\\[-2pt]
Allier	&	438.90	&	320.08	&	-21.08	&	0.76	&	Jura	&	418.98	&	269.47	&	-37.23	&	0.90	\\[-2pt]
Alpes-de-Haute-Provence	&	415.39	&	322.28	&	-22.54	&	0.00	&	Landes	&	1401.96	&	962.94	&	-112.01	&	0.88	\\[-2pt]
Alpes-Maritimes	&	10190.83	&	9766.28	&	4.94	&	0.80	&	Loire	&	312.68	&	160.78	&	-35.28	&	0.83	\\[-2pt]
Ardèche	&	930.77	&	789.08	&	-46.61	&	0.19	&	Loire-Atlantique	&	5291.71	&	4115.61	&	-318.16	&	0.68	\\[-2pt]
Ardennes	&	912.81	&	661.02	&	-24.88	&	0.08	&	Loiret	&	11911.55	&	9882.78	&	-376.13	&	0.38	\\[-2pt]
Ariège	&	323.87	&	256.95	&	-7.27	&	0.73	&	Loir-et-Cher	&	5273.89	&	3967.86	&	-166.97	&	0.83	\\[-2pt]
Aube	&	1133.62	&	1014.17	&	-34.47	&	0.74	&	Lot	&	87.47	&	56.02	&	-7.74	&	0.00	\\[-2pt]
Aude	&	2597.59	&	2073.21	&	-150.56	&	0.68	&	Lot-et-Garonne	&	899.78	&	848.55	&	1.15	&	0.60	\\[-2pt]
Aveyron	&	585.55	&	303.29	&	-59.43	&	0.33	&	Lozère	&	70.27	&	41.71	&	-7.06	&	0.65	\\[-2pt]
Bas-Rhin	&	12961.73	&	9881.53	&	-366.47	&	0.14	&	Maine-et-Loire	&	1031.20	&	863.47	&	-35.49	&	0.62	\\[-2pt]
Bouches-du-Rhône	&	52088.54	&	50810.79	&	-203.33	&	0.78	&	Manche	&	998.81	&	751.93	&	-46.35	&	0.05	\\[-2pt]
Calvados	&	3083.50	&	1902.70	&	-292.93	&	0.71	&	Marne	&	4420.93	&	3940.98	&	37.72	&	0.73	\\[-2pt]
Cantal	&	151.36	&	77.60	&	-16.93	&	0.50	&	Mayenne	&	252.80	&	172.16	&	-16.84	&	0.23	\\[-2pt]
Charente	&	529.13	&	396.29	&	-20.03	&	0.92	&	Meurthe-et-Moselle	&	6388.98	&	4748.65	&	-150.36	&	0.07	\\[-2pt]
Charente-Maritime	&	2321.01	&	890.01	&	-403.46	&	0.13	&	Meuse	&	223.52	&	177.09	&	-2.67	&	0.92	\\[-2pt]
Cher	&	1411.05	&	1129.37	&	-21.34	&	0.86	&	Morbihan	&	1646.36	&	1094.67	&	-155.83	&	0.07	\\[-2pt]
Corrèze	&	224.20	&	156.91	&	-15.65	&	0.03	&	Moselle	&	22823.92	&	18268.69	&	-308.01	&	0.93	\\[-2pt]
Corse-du-Sud	&	205.06	&	195.38	&	-1.15	&	0.80	&	Nièvre	&	188.18	&	109.73	&	-23.32	&	0.14	\\[-2pt]
Côte-d'Or	&	2745.03	&	2113.74	&	-148.13	&	0.84	&	Nord	&	98379.10	&	91468.36	&	-1059.30	&	0.09	\\[-2pt]
Côtes-d'Armor	&	340.92	&	201.88	&	-37.70	&	0.58	&	Oise	&	9708.87	&	7952.05	&	-146.81	&	0.52	\\[-2pt]
Creuse	&	23.74	&	10.50	&	-2.18	&	0.10	&	Orne	&	272.68	&	163.80	&	-17.93	&	0.64	\\[-2pt]
Deux-Sèvres	&	283.06	&	227.86	&	-4.40	&	0.58	&	Paris	&	1202.76	&	1086.63	&	-20.87	&	0.03	\\[-2pt]
Dordogne	&	690.92	&	469.95	&	-42.92	&	0.74	&	Pas-de-Calais	&	32943.70	&	30450.81	&	-190.75	&	0.81	\\[-2pt]
Doubs	&	2869.10	&	2019.97	&	-183.07	&	0.87	&	Puy-de-Dôme	&	2377.93	&	1694.43	&	-180.59	&	0.43	\\[-2pt]
Drôme	&	3680.50	&	3287.95	&	-145.61	&	0.43	&	Pyrénées-Atlantiques	&	3790.71	&	2866.96	&	-143.04	&	0.40	\\[-2pt]
Essonne	&	14145.72	&	12674.88	&	-222.78	&	0.67	&	Pyrénées-Orientales	&	6296.51	&	5771.25	&	-129.40	&	0.49	\\[-2pt]
Eure	&	2667.30	&	1878.33	&	-136.20	&	0.00	&	Rhône	&	28022.35	&	24199.68	&	-738.39	&	0.61	\\[-2pt]
Eure-et-Loir	&	2042.22	&	1660.13	&	1.83	&	0.23	&	Saône-et-Loire	&	2068.73	&	1450.22	&	-123.79	&	0.81	\\[-2pt]
Finistère	&	1860.86	&	1582.40	&	-43.84	&	0.34	&	Sarthe	&	977.44	&	737.69	&	-58.48	&	0.75	\\[-2pt]
Gard	&	9178.11	&	8039.79	&	-197.78	&	0.65	&	Savoie	&	4206.58	&	3206.06	&	-269.02	&	0.00	\\[-2pt]
Gers	&	191.32	&	91.98	&	-18.68	&	0.31	&	Seine-et-Marne	&	27000.56	&	24988.77	&	24.41	&	0.31	\\[-2pt]
Gironde	&	24971.19	&	18396.93	&	-956.52	&	0.94	&	Seine-Maritime	&	19604.19	&	15832.05	&	-490.19	&	0.05	\\[-2pt]
Haute-Corse	&	8116.29	&	5983.59	&	-503.88	&	0.22	&	Seine-Saint-Denis	&	4786.20	&	4575.04	&	20.23	&	0.06	\\[-2pt]
Haute-Garonne	&	605.33	&	493.39	&	-13.07	&	0.57	&	Somme	&	3009.85	&	2556.21	&	-46.82	&	0.52	\\[-2pt]
Haute-Loire	&	21450.80	&	16759.65	&	-1031.05	&	0.87	&	Tarn	&	972.98	&	662.13	&	-56.59	&	0.28	\\[-2pt]
Haute-Marne	&	178.95	&	80.22	&	-34.26	&	0.14	&	Tarn-et-Garonne	&	544.43	&	416.16	&	-21.02	&	0.96	\\[-2pt]
Hautes-Alpes	&	264.15	&	215.94	&	-4.36	&	0.94	&	Territoire de Belfort	&	180.69	&	121.07	&	-16.90	&	0.04	\\[-2pt]
Haute-Saône	&	287.07	&	135.12	&	-37.61	&	0.65	&	Val-de-Marne	&	10839.53	&	10064.38	&	-63.43	&	0.04	\\[-2pt]
Haute-Savoie	&	9050.93	&	7192.27	&	-555.01	&	0.74	&	Val-d'Oise	&	4110.93	&	3859.22	&	-14.53	&	0.31	\\[-2pt]
Hautes-Pyrénées	&	1297.49	&	749.23	&	-95.74	&	0.66	&	Var	&	16036.51	&	15974.75	&	-144.09	&	0.79	\\[-2pt]
Haute-Vienne	&	387.93	&	316.66	&	-23.32	&	0.47	&	Vaucluse	&	4973.49	&	4434.07	&	-161.66	&	0.90	\\[-2pt]
Haut-Rhin	&	550.06	&	428.26	&	-21.10	&	0.00	&	Vendée	&	1528.86	&	1003.22	&	-151.83	&	0.61	\\[-2pt]
Hauts-de-Seine	&	2439.51	&	2317.14	&	2.03	&	0.29	&	Vienne	&	552.11	&	389.17	&	-33.00	&	0.89	\\[-2pt]
Hérault	&	21088.38	&	19014.92	&	-385.19	&	0.01	&	Vosges	&	1162.54	&	673.37	&	-131.30	&	0.04	\\[-2pt]
Ille-et-Vilaine	&	4920.71	&	3332.20	&	-31.54	&	0.93	&	Yonne	&	825.48	&	686.72	&	-7.81	&	0.21	\\[-2pt]
Indre	&	237.86	&	150.06	&	-23.17	&	0.13	&	Yvelines	&	19797.90	&	17327.33	&	-316.48	&	0.23	\\
    \hline
    \end{tabular}
    \end{footnotesize}
    \end{center}
\label{T:result}
\end{table*}

%% file: T2-tab-3.tex
\begin{table*}
    \caption{%
    Yearly avarage of AL for all observatories in France. 
    Column definitions are the same as in Table \ref{T:result}.
    The observatory codes are taken from \cite{2020MNRAS.493.1204A}.
    AL measurements of some observatories were lower than VIIRS radiometric resolution level (3 nW cm$^{-2}$ sr$^{-1}$) and they are marked with dashes. The SQM column values are mpsas units. 
    }
    \begin{center}
    \begin{footnotesize}
    \begin{tabular}{%
        @{}               r@{~}p{33mm}@{~~}r@{~~}r@{~~}r@{~~}r
        @{\hspace*{0.5em}}r@{~}p{45mm}@{~~}r@{~~}r@{~~}r@{~~}r@{}}
    \hline
    Code &
    Obs.
        & \multicolumn{1}{@{}c@{}}{Ave-All}
        & \multicolumn{1}{@{}c@{}}{L.R.}
        & \multicolumn{1}{@{}c@{}}{$R^2$}
        & SQM
        &
    Code &
    Obs.
        & \multicolumn{1}{@{}c@{}}{Ave-All}
        & \multicolumn{1}{@{}c@{}}{L.R.}
        & \multicolumn{1}{@{}c@{}}{$R^2$}
        & SQM
        \\
    \hline
4	&	Montpellier	&	13.78	&	-0.05	&	0.10	&	17.85	&	918	&	Saint-Sulpice	&	-	&	-	&	-	&	-	\\[-2pt]
5	&	Toulouse	&	33.48	&	-0.44	&	0.43	&	17.14	&	920	&	Durtal	&	-	&	-	&	-	&	-	\\[-2pt]
6	&	Meudon	&	13.13	&	0.05	&	0.00	&	17.97	&	928	&	Merignac	&	11.45	&	-1.33	&	0.79	&	18.45	\\[-2pt]
8	&	Paris	&	64.80	&	-1.84	&	0.60	&	16.61	&	929	&	Obs. de Dax	&	4.68	&	-0.10	&	0.16	&	18.76	\\[-2pt]
11	&	Caussols	&	-	&	-	&	-	&	-	&	930	&	Ramonville Saint Agne	&	18.68	&	0.15	&	0.06	&	17.66	\\[-2pt]
15	&	Marseilles	&	69.58	&	-1.06	&	0.70	&	16.54	&	970	&	Bordeaux-Floirac	&	31.36	&	-0.32	&	0.09	&	17.21	\\[-2pt]
17	&	Besancon	&	26.08	&	-1.00	&	0.54	&	17.54	&	975	&	Saint-Caprais	&	-	&	-	&	-	&	-	\\[-2pt]
21	&	Nice	&	16.54	&	0.34	&	0.15	&	17.70	&	976	&	Belesta	&	-	&	-	&	-	&	-	\\[-2pt]
92	&	Obs. de Nurol	&	-	&	-	&	-	&	-	&	978	&	Gretz-Armainvilliers	&	10.18	&	-0.35	&	0.41	&	18.26	\\[-2pt]
123	&	Pises Obs.	&	-	&	-	&	-	&	-	&	979	&	Malibert	&	-	&	-	&	-	&	-	\\[-2pt]
125	&	Castres	&	25.89	&	-0.86	&	0.27	&	17.58	&	980	&	Quincampoix	&	4.76	&	0.00	&	0.00	&	18.74	\\[-2pt]
132	&	Obs. de l'Ardeche	&	-	&	-	&	-	&	-	&	982	&	Wormhout	&	14.02	&	-0.29	&	0.06	&	17.88	\\[-2pt]
133	&	Bedoin	&	-	&	-	&	-	&	-	&	985	&	Les Engarouines Obs.	&	-	&	-	&	-	&	-	\\[-2pt]
134	&	Les Tardieux	&	-	&	-	&	-	&	18.95	&	1029	&	Obs. de Chalandray-Canotiers	&	11.80	&	0.28	&	0.29	&	17.94	\\[-2pt]
139	&	Village-Neuf	&	14.85	&	0.32	&	0.20	&	17.71	&	1034	&	Cosmosoz Obs.	&	22.15	&	0.21	&	0.10	&	17.42	\\[-2pt]
140	&	Antibes	&	28.15	&	0.41	&	0.23	&	17.29	&	1036	&	Le Couvent de Lentin	&	-	&	-	&	-	&	-	\\[-2pt]
141	&	Augerolles	&	-	&	-	&	-	&	-	&	1048	&	Obs. Chante-Perdrix	&	-	&	-	&	-	&	-	\\[-2pt]
142	&	Hottviller	&	-	&	-	&	-	&	-	&	1057	&	Albigneux	&	-	&	-	&	-	&	-	\\[-2pt]
145	&	Bray et Lu	&	4.59	&	-0.22	&	0.43	&	18.77	&	1071	&	Savigny-le-Temple	&	4.80	&	-0.16	&	0.32	&	18.89	\\[-2pt]
149	&	Guitalens	&	4.58	&	-0.12	&	0.16	&	18.90	&	1081	&	Obs. des Baronnies Provencales	&	-	&	-	&	-	&	-	\\[-2pt]
150	&	Beine-Nauroy	&	4.49	&	0.04	&	0.04	&	18.70	&	1095	&	Cesson	&	8.21	&	-0.84	&	0.66	&	18.62	\\[-2pt]
151	&	Maisons Laffitte	&	37.53	&	-0.20	&	0.15	&	17.02	&	1097	&	Obs. des Terres Blanches	&	-	&	-	&	-	&	-	\\[-2pt]
165	&	St. Michel sur Meurthe	&	-	&	-	&	-	&	-	&	1115	&	Eygalayes	&	-	&	-	&	-	&	-	\\[-2pt]
178	&	Le Cres	&	30.11	&	0.36	&	0.48	&	17.17	&	1122	&	Vallauris	&	17.71	&	0.20	&	0.51	&	17.57	\\[-2pt]
179	&	Collonges	&	10.59	&	-0.94	&	0.36	&	18.81	&	1154	&	Gieres	&	13.76	&	0.21	&	0.19	&	17.88	\\[-2pt]
181	&	Mauguio	&	53.83	&	0.12	&	0.01	&	16.73	&	1162	&	Bollwiller	&	8.82	&	-0.07	&	0.05	&	18.30	\\[-2pt]
185	&	Valmeca Obs.	&	-	&	-	&	-	&	-	&	1163	&	Chinon	&	-	&	-	&	-	&	-	\\[-2pt]
200	&	Buthiers	&	-	&	-	&	-	&	-	&	1180	&	Rouet	&	-	&	-	&	-	&	-	\\[-2pt]
203	&	Tamaris Obs.	&	17.90	&	0.26	&	0.47	&	17.58	&	1181	&	Maisoncelles	&	-	&	-	&	-	&	-	\\[-2pt]
217	&	Obs. des Cote de Meuse	&	-	&	-	&	-	&	-	&	1190	&	ROSA Obs.	&	-	&	-	&	-	&	-	\\[-2pt]
223	&	Yerres-Canotiers	&	27.73	&	0.09	&	0.06	&	17.26	&	1230	&	Chelles	&	44.13	&	-0.28	&	0.04	&	17.00	\\[-2pt]
225	&	Ottmarsheim	&	6.52	&	-0.19	&	0.18	&	18.38	&	1264	&	SATINO Remote Obs.	&	-	&	-	&	-	&	-	\\[-2pt]
238	&	Baugy	&	6.08	&	0.17	&	0.40	&	18.47	&	1603	&	Guernanderf	&	-	&	-	&	-	&	-	\\[-2pt]
243	&	Varennes	&	-	&	-	&	-	&	-	&	1616	&	Salvia Obs., Saulges	&	-	&	-	&	-	&	-	\\[-2pt]
273	&	Nimes	&	112.34	&	0.75	&	0.08	&	16.10	&	1636	&	St Pardon de Conques	&	-	&	-	&	-	&	-	\\[-2pt]
276	&	Flammarion Obs.	&	42.35	&	-1.08	&	0.65	&	16.99	&	1666	&	Centre Astro. de La Couyere	&	-	&	-	&	-	&	-	\\[-2pt]
443	&	Merlette	&	16.95	&	0.05	&	0.00	&	17.59	&	1723	&	Sainte Helene	&	5.10	&	-0.56	&	0.87	&	18.93	\\[-2pt]
449	&	Lamalou-les-Bains	&	5.37	&	0.22	&	0.50	&	18.51	&	1749	&	Obs. de Gravelle, St. Maurice	&	53.15	&	-0.81	&	0.27	&	16.83	\\[-2pt]
450	&	Sollies-Pont	&	21.89	&	-1.72	&	0.47	&	17.62	&	1752	&	Micro Palomar, Reilhanette	&	-	&	-	&	-	&	-	\\[-2pt]
475	&	Le Creusot	&	-	&	-	&	-	&	-	&	1753	&	Obs. de Pommier	&	-	&	-	&	-	&	-	\\[-2pt]
480	&	La Seyne sur Mer	&	15.37	&	0.21	&	0.42	&	17.74	&	1757	&	Murviel-les-Montpellier	&	-	&	-	&	-	&	-	\\[-2pt]
482	&	Haute Provence	&	-	&	-	&	-	&	-	&	1758	&	Reilhanette	&	-	&	-	&	-	&	-	\\[-2pt]
484	&	Lyons	&	20.19	&	0.03	&	0.00	&	17.60	&	1760	&	Hesingue	&	9.00	&	0.04	&	0.01	&	18.01	\\[-2pt]
493	&	Strasbourg	&	47.79	&	-0.40	&	0.04	&	16.87	&	1761	&	PASTIS Obs., Banon	&	-	&	-	&	-	&	-	\\[-2pt]
557	&	Pic du Midi	&	-	&	-	&	-	&	-	&	1763	&	Saint-Saturnin-les-Avignon	&	19.16	&	-0.20	&	0.19	&	17.60	\\[-2pt]
585	&	Soisy-sur-Seine	&	16.46	&	0.21	&	0.17	&	17.68	&	1764	&	Les Barres Obs., Lamanon	&	4.32	&	0.02	&	0.05	&	18.81	\\[-2pt]
586	&	St. Veran	&	-	&	-	&	-	&	-	&	1767	&	Saint-Michel-l'Obs.	&	-	&	-	&	-	&	-	\\[-2pt]
588	&	Arbonne la Foret	&	6.94	&	0.30	&	0.70	&	18.25	&	1769	&	St-Martin Obs.	&	-	&	-	&	-	&	-	\\[-2pt]
589	&	Martigues	&	18.81	&	-0.40	&	0.43	&	17.60	&	1839	&	Freconrupt	&	-	&	-	&	-	&	-	\\[-2pt]
598	&	Blauvac	&	-	&	-	&	-	&	-	&	1856	&	Plntrm. de Vaulx-en-Velin Obs.	&	57.05	&	1.37	&	0.21	&	16.62	\\[-2pt]
601	&	Osenbach	&	-	&	-	&	-	&	-	&	1869	&	Premote Obs.	&	-	&	-	&	-	&	-	\\[-2pt]
605	&	Crolles	&	-	&	-	&	-	&	-	&	2077	&	Landehen	&	-	&	-	&	-	&	-	\\[-2pt]
606	&	Pergignan	&	103.17	&	-1.36	&	0.24	&	16.27	&	2131	&	OPERA Obs.	&	-	&	-	&	-	&	-	\\[-2pt]
881	&	Caussols-ODAS	&	-	&	-	&	-	&	-	&		&		&		&		&		&		\\
    \hline
    \end{tabular}
    \end{footnotesize}
    \end{center}
\label{T:obs}
\end{table*}